\documentclass[reprint, nofootinbib, aps, superscriptaddress, prd]{revtex4-2}

\usepackage{inputenc}
\usepackage{bm}
\usepackage{amsmath}
\usepackage{amsfonts}
\usepackage{amssymb}
\usepackage{mathptmx}
\usepackage{graphicx}
\usepackage{color}
\RequirePackage[colorlinks,citecolor=darkred,urlcolor=magenta,linkcolor=darkblue,bookmarks=false]{hyperref}
\usepackage{adjustbox}
\usepackage{comment}
\usepackage{multirow}
\usepackage{makecell}
\usepackage{csquotes}

\usepackage[normalem]{ulem}

\allowdisplaybreaks
\renewcommand\apj{{\rm ApJ}}
\newcommand\apjl{{\rm ApJL}}     
\newcommand\aap{{\rm A\&A}}
\newcommand\aaps{{\rm A\&AS}}
\newcommand\mnras{{\rm MNRAS}}
\renewcommand\prl{{\rm PhRvL}}

\newcommand\jcap{{\rm JCAP}}

\newcommand{\healpix}{\ensuremath{\tt HEALPix}}
\newcommand{\NILC}{\ensuremath{\tt NILC}}

\newcommand{\COrE}{\ensuremath{\tt COrE }}
\newcommand{\camb}{\ensuremath{\tt CAMB}}
\newcommand{\planck}{{\it Planck}}
\newcommand{\litebird}{\ensuremath{\tt LiteBIRD}}

\newcommand{\lenspix}{\ensuremath{\tt LensPix }}
\newcommand{\GNILC}{\ensuremath{\tt GNILC }}

\newcommand{\GNILCn}{\ensuremath{\tt GNILC-dust }}
\newcommand{\GINES}{\ensuremath{\tt MKD-dust }}
\newcommand{\HDdust}{\ensuremath{\tt HD-dust }}

\newcommand{\powerlaw}{\ensuremath{\tt power-law }}

\newcommand{\ECHO}{\ensuremath{\tt ECHO}}
\newcommand{\pico}{\ensuremath{\tt PICO}}

\newcommand{\cmbbh}{\ensuremath{\tt CMB-BHARAT}}

\newcommand{\Nside}{\ensuremath{N_{\rm side}}}

\def\GHz{\ifmmode $\,GHz$\else \,GHz\fi}
\def\muK{\ifmmode \,\mu$K$\else \,$\mu$\hbox{K}\fi}
\def\MJysrmK{\ifmmode \,$MJy\,sr\mo$\,mK$_{\rm CMB}\mo\else \,MJy\,sr\mo\,mK$_{\rm CMB}\mo$\fi}
\newcommand{\daa}{\textcolor{cyan}} 

\newcommand\subsubsubsubsection{\@startsection{subparagraph}{5}{\z@}{-2.5ex\@plus -1ex \@minus -.25ex}{1.25ex \@plus .25ex}{\normalfont\normalsize\bfseries}}
\makeatother

\definecolor{darkblue}{rgb}{0.0, 0.0, 0.62}
\definecolor{deepmagenta}{rgb}{0.8, 0.0, 0.7}
\definecolor{darkred}{rgb}{0.55, 0.0, 0.0}
\definecolor{violet}{rgb}{0.56, 0.0, 1.0}
\newcommand{\reff}{\textcolor{black}}
\newcommand{\rev}{\textcolor{black}}

\begin{document}

\title{Importance of high-frequency bands for removal of thermal dust in \ECHO}
\author{Aparajita Sen}
\email{aparajita15@iisertvm.ac.in}
\affiliation{School of Physics, Indian Institute of Science Education and Research Thiruvananthapuram, Thiruvananthapuram 695551, India}
\author{Soumen Basak}
\affiliation{School of Physics, Indian Institute of Science Education and Research Thiruvananthapuram, Thiruvananthapuram 695551, India}
\author{Tuhin Ghosh}
\affiliation{National Institute of Science Education and Research, An OCC of Homi Bhabha National Institute, Bhubaneswar 752050, Odisha, India}
\author{Debabrata Adak}
\affiliation{The Institute of Mathematical Sciences, CIT Campus, Tharamani, Chennai, Tamil Nadu 600113, India}
\author{Srijita Sinha}
\affiliation{National Institute of Science Education and Research, An OCC of Homi Bhabha National Institute, Bhubaneswar 752050, Odisha, India}


\begin{abstract}
The Indian Consortium of Cosmologists has proposed a cosmic microwave background (CMB) space mission, Exploring Cosmic History and Origin (\ECHO). A major scientific goal of the mission is to detect the primordial $B$-mode signal of CMB polarization. The detection of the targeted signal is very challenging as it is deeply buried under the dominant astrophysical foreground emissions of the thermal dust and the Galactic synchrotron. To facilitate the adequate subtraction of thermal dust, the instrument design of \ECHO\ has included nine dust-dominated high-frequency bands over the frequency range 220-850 GHz. In this work, we closely reexamine the utility of the high-frequency \ECHO\ bands in foreground subtraction using the Needlet Internal Linear Combination component separation method. We consider three dust models: a physical dust model, a dust spectral energy distribution (SED) with a single modified black body (MBB) emission law and a multilayer dust model with frequency-frequency decorrelation. We consider eleven \ECHO\ bands in the $28-190$ \GHz\ range as our baseline configuration \rev{and investigate the changes in the level foreground and noise residuals} as subsequent dust-dominated high-frequency bands are added. We find that adding the high-frequency bands leads to a consistent decrease in the level of residual foreground and noise, and the sensitivity of $r$ measurement improves. Most of the reduction in both residual levels and enhancement in the sensitivity is achieved in the $28-600$\GHz\ frequency range. Negligible change in residual levels is seen by extending the frequency range from $600$\GHz\ to $850$\GHz. We find that extending the \ECHO\ frequency bands from $190$\GHz\ to $340$\GHz\ leads to a $40\%-50\%$ reduction in the foreground and noise residual levels in the recovered CMB map. Correspondingly the sensitivity of \ECHO\ toward $r$ also improves by a similar amount. Furthermore, incorporating higher frequencies up to $600$\GHz\ yields an additional reduction of $12\%-15\%$ in the residual levels and uncertainty on $r$. However, extending observations up to the $850$\GHz\ frequency band only leads to a marginal improvement in sensitivity, ranging from $3\%-7\%$.

\end{abstract}

\maketitle
\section{Introduction}
\label{sec:intro}

The inflationary paradigm, the scenario that the Universe underwent an era of exponential expansion almost at the beginning of its history, is currently the most acceptable model to describe the evolution of the early Universe and the generation of primordial fluctuations which seeded the large-scale structures that we see today \citep{STAROBINSKY:1980,1980ApJ...241L..59K,Guth:1980,10.1093/mnras/195.3.467,linde1982new,PhysRevLett.48.1220}. One of the robust predictions of the inflationary models is the existence of 
primordial $B$-modes, a swirling polarization pattern in the CMB \citep{PhysRevLett.78.2058,PhysRevD.55.1830}. This particular signal is sourced by the primordial gravitational waves during the epoch of inflation. The amplitude of the primordial CMB $B$-modes power spectrum is parameterized by $r$, which is directly related to the energy scale of the inflaton \citep{doi:10.1146/annurev-astro-081915-023433}. Potential detection of the \reff{primordial CMB $B$-modes} will open up a new window into the physics of the early Universe. The recent analysis of BICEP/Keck data combined with the Planck and WMAP data has put an upper limit of $r<0.036$ at $95\%$ confidence level \citep{tristram2021planck_BK18}. The next generation of CMB experiments is expected to improve on this upper limit by an order of magnitude or so.

Quite a few experimental and theoretical issues are currently under investigation in order to assess the feasibility of the detection of the primordial CMB $B$-modes signal. The signal is expected to be very weak compared to the CMB temperature anisotropies and $E$-modes of polarization, and most of its power is expected to be concentrated at low multipoles (or large angular scales). The main problematic aspects come from the polarization of astrophysical foregrounds, instrumental noise and systematic effects. These effects are expected to mask the signal from the primordial tensor modes. The gravitational lensing on the CMB by the matter distribution adds to these complications. 
Detection of the faint $B$-mode signal will require not only extremely sensitive instruments to observe the sky from space but also exquisite control over systematic errors. Furthermore, the signal is deeply buried under the astrophysical foregrounds by several orders of magnitude. Its extraction can only be achieved by efficient component separation techniques \citep{Eriksen:2004, Basak_and_Delabrouille:2012, Remazeilles:2020}.

The component separation methods utilize, in general, the distinct spectral properties of cosmological and astrophysical components to disentangle the CMB and other astrophysical components \citep{10.1093/ptep/ptu065}. One of the ways to improve the efficiency of the component separation technique is to obtain sky observations for a broad range of frequency coverage where the observed sky at the low and high-frequency range, respectively, play the role of tracers of synchrotron emission and thermal dust emission. Keeping these requirements in mind, an Indian Consortium of Cosmologists has proposed a space mission named Exploring Cosmic History and Origin (\ECHO), popularly known as \cmbbh\footnote{\url{http://cmb-bharat.in/}}. Other endeavours for space based experiments with similar objectives include \COrE\ \citep{core_instrument:2018},  \pico\ \citep{PICO:2019} and \litebird\ \citep{Hazumi:2019}. 


\litebird, \ECHO\ and  \COrE\ aim to \reff{constrain $r\leq  10^{-3}$ at} $3\sigma$ detection level while \pico\ which is more ambitious will constrain $r\leq 2.7\times 10^{-4}$ with  95\% confidence level 
\citep{aurlien2022foreground}. While \litebird\ has been designed primarily to detect the primordial CMB $B$-mode signal, the other space missions have a broad range of scientific goals to achieve. So, \litebird\ will focus on the measurement of the sky mostly at large angular scales and will be equipped with $15$ frequency bands spread over the range of 34-448 GHz \citep{Hazumi:2019}. On the other hand, \ECHO\ will observe the sky at $20$ frequency bands over a wider frequency range of 28-850 GHz, at a beam resolution ranging from $40'$ to $1.3'$. The high-resolution sky measurements from \ECHO\ 
will be used to put stringent constrain on the sum of neutrino masses and improve the uncertainty over the measurement of the scalar spectral index ($n_s$). The high-frequency bands ($\geq 353$\,\GHz)\ of \ECHO\ will be used to obtain reliable maps of the cosmic infrared background (CIB) emission and to constraint the components of the interstellar dust and the Galactic magnetic field. Similarly, \pico\ will observe microwave sky at $21$ frequency bands in the range of 20-800 GHz with a beam resolution ranging from $38'$ to $1'$.

Recovering $r$ as low as $10^{-3}$ through component separation is an extremely challenging task \citep{Katayama:2011,10.1093/mnras/stw441}. At frequencies $\ge70$ GHz, the CMB foreground is mostly dominated by thermal dust emission. The analysis done by the authors in \cite{Hensley_2018} confirms that the recovery of the CMB $B$-mode signal after the component separation can be severely biased by the dust complexities. They have explored the optimum configuration of frequency bands required to minimize foreground residuals from thermal dust emission. Furthermore, they have reported that for certain forms of dust, it is better to limit our observations at lower frequency bands of $\approx 200-500$ \GHz.\ The addition of higher frequency channels in the range of 600-800 GHz can significantly bias the CMB measurement. However, this study is restricted to a single pixel of the sky where it is less polluted, using a parametric method for component separation. This paper is an extension of \cite{Hensley_2018} work where we do a similar exercise using a blind component separation method called Needlet Internal Linear Combination (\NILC) over a significant fraction of the sky.

The main goal of our work \reff{is to explore the utility of the high-frequency bands of \ECHO\ for the effective removal of thermal dust contamination.} There are $9$ frequency channels between $220$ and $850$ GHz that will act as tracers of the thermal dust emission. 
This instrument design is made under the assumption that the observation of dust SED over a wide range of frequencies will help in breaking the degeneracy among the multiple dust parameters and increase the efficiency of the component separation techniques. We test this assumption by dropping the high-frequency channels of the \ECHO\ mission to extract the CMB $B$-mode signal with minimal foreground contamination. 
Unlike the single-pixel analysis of \cite{Hensley_2018}, we use $30\%-60\%$ of the unmasked sky to recover the CMB $B$-mode signal using \NILC\  \citep{Basak_and_Delabrouille:2012,Basak:2013}. \reff{The \NILC\ pipeline used in our work has been tested on simulations of the sky for Planck and extensively used for the analysis of Planck data ~\citep{diffuse_planck_2013,planck-IV:2018}. Following this work, the same pipeline has also been used to forecast the tensor-to-scalar ratio in \litebird\ ~\citep{forecast-litebird}, \ECHO\ ~\citep{10.1093/mnras/stac1474} and \COrE\ ~\citep{ CORE-B:2018}}. Here, we make forecasts on $r$ and its standard deviation ($\sigma_r$) derived from the recovered CMB $B$-mode maps for various combinations of high-frequency channels for the \ECHO\ mission.

The paper is arranged as follows. In Sec.~\ref{sec:freq_channels}, we describe the instrumental configuration of \ECHO\ briefly. Next, we describe the various models of CMB and foregrounds used for simulating the microwave sky as seen by \ECHO\ in Sec.~\ref{sec:Sky Sim}. This is followed by a discussion about the procedure used to extract $r$ from the realistic sky simulations in Sec.~\ref{sec:methodology}. Finally, we present our results in Sec.~\ref{sec:results} and conclude in Sec.~\ref{sec:Conclusions}.

\section{Frequency Channels of \ECHO}
\label{sec:freq_channels}
\begin{figure*}
    \includegraphics[scale=0.27]{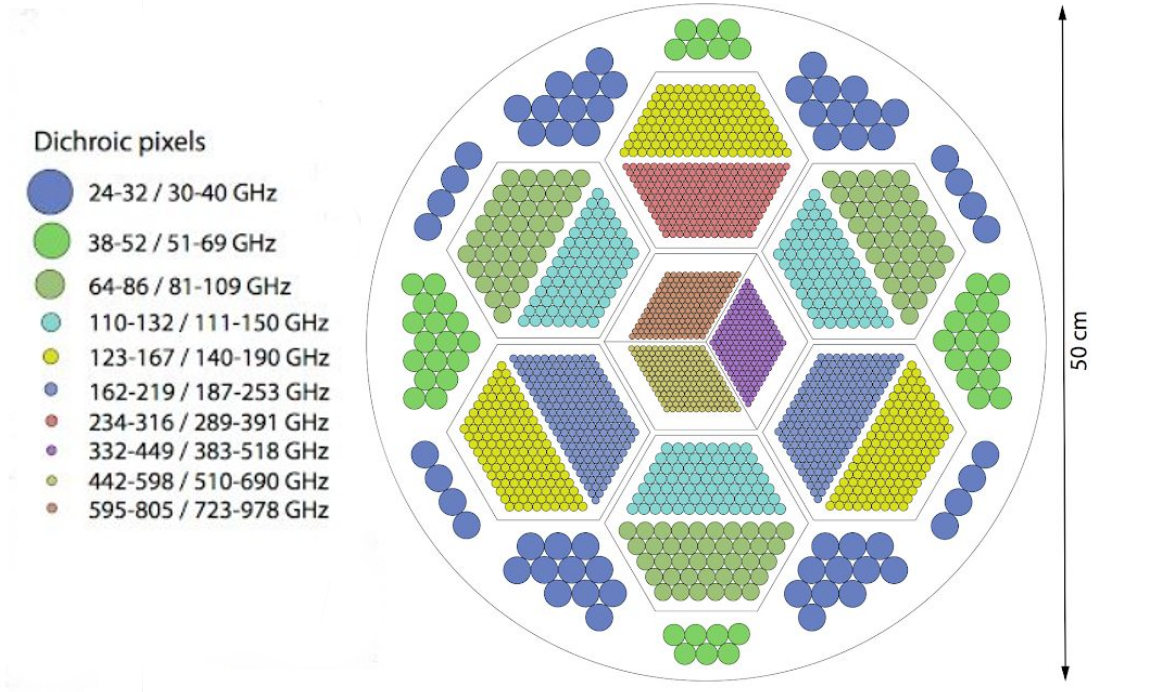}
    \caption{A schematic diagram representing the \ECHO\ focal plane. The focal plane will consist of about 8000 detectors/pixels mounted over 7 wafers. The colored circles represent the horns at different frequency bands.}
    \label{fig:echo_fpa}
\end{figure*}

\rev{The quest for primordial CMB $B$-modes} will require designing extremely sensitive detectors. Due to recent technological advancements, the sensitivity of the detectors is close to the photon noise limit. In this scenario, the sensitivity of the instrument is further improved by increasing its optical throughput. This has become possible due to the advent of new detector technologies, which allow for the installation of up to $10^{3}-10^{5}$ pixels with bolometric detectors in a small focal plane area (FPA). Furthermore, the use of multichroic pixels, which are simultaneously sensitive at different frequencies allows for designing compact FPAs sensitive to a broad range of frequencies. 
\reff{The telescope installed in \ECHO\ will have a diameter of $1.5$ meters} consisting of a focal plane of diameter $52$ cm. The focal plane will be equipped with an array of approximately $8000$ polarization-sensitive dichroic pixels, fabricated on seven hexagonal wafers. \ECHO\ will observe the microwave sky through $20$ frequency bands with an approximate bandwidth of $30\%$. Fig.~\ref{fig:echo_fpa} shows a schematic representation of the instrument's focal plane. The colored circles in the figure signify the two frequency bandwidth at which the pixels are sensitive. The high-frequency detectors will be placed in the centre of the focal plane, while the larger aperture detectors for lower frequency will be installed in the periphery of the focal plane. The effective sensitivity $s$ of each frequency band depends on several factors ~\citep{Wu_2014, 10.1111/j.1365-2966.2004.07506.x};      
\begin{equation}
    s^i [\mu K.arcmin] \equiv \frac{NET^i[\mu K.\sqrt{s}]\times \sqrt{f_{sky}[arcmin^{2}]}}{\sqrt{N_{det}^i\times \Delta T[s]}}.
\end{equation}

Here, $NET$ is the noise equivalent temperature of the detectors. It is defined as the amount of signal required from the source in an interval of one second to get a signal-to-noise ratio of one. $f_{sky}$ stands for the fraction of sky surveyed and $\Delta T$ is the observational period for \ECHO\ (around 4 years). $N_{det}^{i}$ is the total number of detectors for the $i$th frequency band. The most sensitive frequency channels of \ECHO\ (around 145 GHz) will have more than 150 detectors for each band. The number of detectors and the polarization sensitivity $s^{i}$ for the \ECHO\ frequency channels are listed in Table~\ref{table:r.m.s.}.

The instrumental design of the \ECHO\ is still in its proposal phase, which means that there is a lot of scope for further improvement and optimization. Upon establishing that higher frequency channels are redundant for the measurement of the primordial CMB $B$-mode signal, we can modify the design of the instrument to our advantage. For instance, FPA can be reduced, or more detectors can be added at other crucial frequency bands (keeping FPA fixed) to improve the overall instrument sensitivity. 
We could also add new low-frequency bands below $28$ GHz in the FPA. The lower frequency bands can help in breaking the degeneracy between the Anomalous Microwave Emission (AME) and the synchrotron \cite{10.1093/mnras/stw441}. Additionally, these bands will enable more accurate characterization of the complexities in synchrotron emissions and thus a better performance of the component separation method ~\citep{10.1093/mnras/stac1474}. 

\begin{table}
\caption{Specifications of the \ECHO\ instrument as proposed by the CMB-Bharat consortium. The instrument will have around $8000$ detectors in its FPA distributed over the 20 frequency bands from 28-850 GHz .}
    \begin{tabular}{ p{2 cm}  p{1.5cm}  p{1.2cm}  p{1.5cm}  }
    \hline \hline \noalign{\vskip 2pt}
    Frequency Band Centers & Beam FWHM  & $N_{DET}$ & Sensitivity $s$\\
   (GHz)&(arcmin)&&($\mu$K.arcmin)\\
   \hline \noalign{\vskip 2pt}
   28&39.9&120&16.5\\
   35&31.9&120&13.3\\
   45&24.8&96&11.9\\
   65&17.1&96&8.9\\
   75&14.91&240&5.1\\
   95&11.7&240&4.6\\
   115&9.72&462&3.1\\
   130&8.59&462&3.1\\
   145&7.70&810&2.4\\
   165&6.77&810&2.5\\
   190&5.88&752&2.8\\
   220&5.08&752&3.3\\
   275&4.06&444&6.3\\
   340&3.28&444&11.4\\
   390&2.86&338&21.9\\
   450&2.48&338&43.4\\
   520&2.14&338&102.0\\
   600&1.86&338&288.0\\
   700&1.59&338&1122.0\\
   850&1.31&338&9550.0\\
    \hline
    \hline
    \end{tabular}
    \label{table:r.m.s.}
\end{table}
\section{Sky simulations}
\label{sec:Sky Sim}

In this section, we describe the foreground models used to simulate the microwave sky as seen by \ECHO\ and its instrumental noise. To simulate these components we use the \ensuremath{PySM} \citep{thorne2017python} and \ensuremath{PSM} \citep{delabrouille2013pre} packages. 
In the \ECHO\ frequency channels, the CMB signal is expected to be buried under the foreground emissions, such as thermal dust, synchrotron, AME, and point sources. However, it has been pointed out very recently in a companion paper \citep{10.1093/mnras/stac1474} on \ECHO,\ that the presence of AME does not affect the recovery of $r$ while the unresolved radio and infrared  point sources introduce only a small bias. Therefore, we consider only the dominant polarized foreground components, thermal dust and synchrotron in the current work. Here, we choose lensed CMB maps with null input tensor-to-scalar ratio as our input CMB maps (see Sec.~\ref{sec:cmb}). To simulate the contributions from synchrotron emissions, we use the standard \powerlaw\ model (described in Sec.~\ref{sec:synchrotron}). Since our analysis is focused on the removal of foregrounds dominated at high frequencies of \ECHO, we consider three models of thermal dust emission with different complexities to make our study robust. These models have been described in Sec.~\ref{sec:thermaldust}.  We simulate the Gaussian instrumental noise based on the projected sensitivity levels described in Table ~\ref{table:r.m.s.}. Finally, we coadd the simulations of the CMB, synchrotron, thermal dust, and noise to obtain three different models for the \ECHO.\  In Table~\ref{table:model-description}, we have listed the name of these simulations and the corresponding foreground models used. All the simulated maps have been generated on \healpix\ grid at a resolution of \Nside$ = 256$ followed by smoothing with a Gaussian beam of FWHM = $40'$.

\begin{table*}
\caption{Description of the set of sky simulations used in our analysis. The sky simulation consists of the CMB, thermal dust, and synchrotron components. The table below lists the name assigned to each set of simulations, and the modeling used for the thermal dust and synchrotron  emissions respectively.}
\label{table:model-description}
   \begin{centering}
   \begin{tabular}{ p{0.3\linewidth}  p{0.3\linewidth}  p{0.3\linewidth}}
   \hline
   \hline \noalign{\vskip 2pt}
    Sky Model & Thermal dust & Synchrotron\\
    \hline
    $M1$ & \HDdust & power-law\\
    $M2$ & \GNILCn & power-law\\
    $M3$ & \GINES & power-law\\
    \hline
    \hline
\end{tabular}
\end{centering}

\end{table*}

\subsection{CMB}
\label{sec:cmb}
 
We use the \lenspix\footnote{\url{http://cosmologist.info/lenspix/}} package to simulate the lensed CMB $B$-modes. For our analysis, we set the input tensor-to-scalar ratio $r_{in}=0$. The theoretical angular power spectra and the lensing potential are used as inputs in the \lenspix\ package and have been generated using \camb\footnote{\url{https://camb.info/}} for \planck\ 2018 best-fit $\Lambda$CDM  parameters ~\citep{planck2018_param}. 

\subsection{Synchrotron model}
\label{sec:synchrotron}
Synchrotron emission is non-thermal radiation emitted by relativistic cosmic ray electrons gyrating around the Galactic magnetic field (GMF). 
The directional nature of the synchrotron emission perpendicular to the ambient magnetic field makes it highly polarized. %
The SED of synchrotron emission depends on the energy distribution of relativistic cosmic ray electrons. In this paper, we assume a power-law SED for the synchrotron emission,

\begin{align}
    &Q^{s}_{\nu} \, =\, Q^{s}_{\nu_0} \left (\frac{\nu}{\nu_0 } \right )^{\beta_{s}}, \nonumber\\
    &U^{s}_{\nu} \, =\, U^{s}_{\nu_0} \left (\frac{\nu}{\nu_0} \right )^{\beta_{s}}. 
\end{align}
We simulate synchrotron frequency maps using the \ensuremath{PySM} $s1$ model. 
The $9$-year WMAP polarization data \citep{bennett2013nine} at  23 \GHz\ smoothed with $3$ degrees Gaussian beam is set as template Stokes parameters, $Q^{s}_{\nu_0}$ and $U^{s}_{\nu_0}$ at the reference frequency $\nu_0 =$ 23 \GHz. The small-scale features are added to the template maps using the methodology described in ~\cite{thorne2017python}. The spectral index map, $\beta_{s}$ used in this model varies around its mean value of $-3.0$. This map is  generated based on model-4 of  \cite{Miville-Desch:2008}  and employing  WMAP $23$ GHz polarization map and 408 MHz Haslam map \citep{Haslam:1982}. 


\subsection{Thermal dust}
\label{sec:thermaldust}
For frequencies above 70 GHz, polarized emission from dust grains in our galaxy is the dominant foreground contamination for CMB polarization. The elongated dust grains align themselves with ambient GMF and preferentially emit radiation parallel to their longer axis, which leads to dust emission being polarized. Current Planck observations suggest that the thermal dust emission can be polarized up to $20\%-25\%$ at high Galactic latitudes \citep{planck-XI:2018}.  Thermal dust emission is empirically well described by a single MBB (or 1MBB) spectrum in thin optical limits at submillimetre bands. 

In practice, the three-dimensional structure of the interstellar medium makes the  physics of thermal dust emission more complex. The dust properties, such as grain sizes, chemical compositions, grain's intrinsic polarization fractions, and local GMF, vary in different regions of our Galaxy. Even if individual dust clouds along the line-of-sight (LOS) follow the MBB spectrum, what we observe is the superposition of many such MBB emissions. Furthermore, multiple dust clouds of different compositions of dust grains have different alignment efficiency to misaligned GMF along a single LOS. This leads to dust polarization angle being  different at different frequencies (frequency-frequency decorrelation \cite{2015MNRAS.451L..90T,2021A&A...647A..16P}.) Therefore, a single MBB (hereafter 1MBB)  spectral model is an approximation to describe all such complex physics of polarized dust emission. Moreover, the averaging over the SEDs along the LOS or the instrumental beam results in a deviation from the 1MBB  description of dust spectrum \citep{chluba2017rethinking} (averaging effect). 

While these complexities of dust polarization modeling are not significant at the sensitivity level of Planck, several studies have shown ~\citep{10.1093/mnras/stw441,10.1093/mnras/stac1474,Sponseller_2022} that they exert a major challenge in the unbiased estimation of $r$. In this work, we consider three dust polarization models. First, the physical dust model accounts for the variation in dust grain composition, next, the standard 1MBB model and lastly, a multilayer dust model, which models the 3D variation of dust spectral properties along the LOS. We have described these models in the following sections. In Appendix~\ref{sec:averaging effect}, we have described the impact of averaging effect in the detection of $r$.  

\subsubsection{$D1$: \HDdust}
\label{ssec:$D1$-Physical}

Depending on the physical properties of the dust grains and the ambient radiation field, the thermal dust SED can be modeled as
\begin{equation}
I_{\nu}=\sum_{j}\int da\frac{dm_{j}}{da}\int dT\left(\frac{dP}{dT}\right)_{\chi,a,j} \tau_{\nu,j,a} B_{\nu}(T) \ ,    
\end{equation}
where $B_{\nu}$ is the \planck\ function, and a specific form of dust composition is indexed by $j$. We denote the radius of the dust grain by $a$ and it's mass with $m_{j}$. The term $dT\left(\frac{dP}{dT}\right)$ denotes the probability with which a dust grain will attain a temperature between $T$ and $T+dT$. This quantity depends on the radiation field, grain size and composition.  The dust opacity ($\tau$) will vary with frequency and dust grain property. It can be derived from the dielectric function of the given grain material. Similarly, we can also model the polarization SED with an additional parameter $f(a,j)$ describing the number of dust grains aligned with the GMF. Based on this physical modeling, in \cite{Hensley_2017}, the authors propose a model in which two forms of dust composition are considered, silicon and carbonaceous. Observations from dust extinction and infrared emission maps are used to constrain the physical properties of the dust grain.

\subsubsection{$D2$: \GNILCn}
\label{ssec:D2-1MBB}
We simulate the most commonly used thermal dust model with 1MBB spectrum using \ensuremath{PSM}. First, the thermal dust intensity maps $I^{\GNILC}_{\nu}$ is computed as follows;
\begin{equation}
    I^{\GNILC}_{\nu} = \tau_{0}\left (\frac{\nu}{\nu_{0}}\right )^{\beta_d} B_{\nu} (T_{d}). 
    \label{eq:D1}
\end{equation}
Here $\tau_{0}$, $\beta_d$ and $T_d$ are the dust opacity, dust spectral index and temperature, respectively, at the reference frequency, $\nu_{0} = 353$ GHz. These parameters vary spatially across the sky and have been estimated by fitting the intensity maps at \planck\ high-frequency channels and the \textit{IRIS} $100\,\mu m$ map. Before fitting, the contribution from the cosmic infrared background (CIB) has been subtracted from observed intensity maps using the \GNILC\ \citep{Remazeilles:2011} method. The $ I^{\GNILC}_{\nu}$ maps are translated to corresponding Stokes $Q$ and $U$ maps as, 
\begin{align}
& Q^{d}_{\nu} = f_d \, g_d \, I^{\GNILC}_{\nu} \, \cos\left(2\gamma_{d}\right),\nonumber\\
& U^{d}_{\nu} = f_d \, g_d \, I^{\GNILC}_{\nu} \, \sin\left(2\gamma_{d}\right).
\label{eq:d2}
\end{align}
Here, $\gamma_{d}$ is the dust polarization angle, $f_d$ and $g_d$ are the dust polarization fraction and the depolarization factors respectively. Both the polarization angle and the depolarization factor depend on the orientation of the GMF along the LOS \citep{Miville-Desch:2008}. The $\gamma_{d}$ and $g_d$ maps are derived at scales larger than 20 degrees from modeling of the GMF in \citep{Miville-Desch:2008} and  23 GHz WMAP and Haslam 408 MHz data. The small scale information of $\gamma_{d}$ and $g_d$ have been added  to the model following \cite{Miville:2007}. The methodology to combine information from different scales has been discussed in detail in ~\citep{delabrouille2013pre}. In this modeling, we take $f_{d}=0.15$, and the polarization fraction is obtained by modulating this quantity with depolarization factor $g_d$.  \reff{On average }$f_d\,g_d \approx 0.05$ with some spatial variation in the sky.

\subsubsection{$D3$: \GINES}
\label{ssec:D3-MKD}
We use a multilayer dust model of \cite{gines:2018} which accounts for the three-dimensional variation of dust emission properties along the LOS as follows,



\begin{align}
\label{eq:MKD-dust} 
& I_{\nu}=\int_{0}^{\infty} dx \frac{d\tau(x,\nu_{0})}{dx}\left(\frac{\nu}{\nu_{0}}\right)^{\beta(x)} B_{\nu}(T(x)), \nonumber\\
& Q_{\nu}=\int_{0}^{\infty} dx p(x) \frac{d\tau(x,\nu_{0})}{dx}\left(\frac{\nu}{\nu_{0}}\right)^{\beta(x)} B_{\nu}(T(x) \cos{2\psi(x)}\sin^{k}{\alpha(x)}, \\
& U_{\nu}=\int_{0}^{\infty} dx p(x) \frac{d\tau(x,\nu_{0})}{dx}\left(\frac{\nu}{\nu_{0}}\right)^{\beta(x)} B_{\nu}(T(x)\sin{2\psi(x)}\sin^{k}{\alpha(x)}.\nonumber
\end{align}

Here, $x$ is the distance along the LOS, $p(x)$ parameterises the local intrinsic polarization fraction of grains and $\psi(x)$ is the polarization angle. $\sin^k{\alpha} (x)$ accounts for alignment efficiency of grains with assumption of $k = 3$ following \cite{Fauvet:2011}. $\psi (x)$ and $\sin^k{\alpha} (x)$ depend upon the orientation of GMF with the LOS.  In \cite{gines:2018}, the authors approximate the continuum emission in Eq.~\ref{eq:MKD-dust} as a discrete sum of six independent layers that follow different emission laws, 
\begin{equation}
    I_{\nu}=\sum_{i=1}^{N} \tau_i(\nu_{0})\left(\frac{\nu}{\nu_{0}}\right)^{\beta_i} B_{\nu}(T_i).
    \label{eq:MKD4}
\end{equation}
Each layer of emission is loosely associated with emission coming from a particular distance from our solar system. Six dust extinction maps of \cite{Green:2015} have been used as proxy of optical depth maps $\tau_i(\nu_{0})$ at reference frequency $\nu_{0}$. 
The spectral maps of $T_i$ and $\beta_i$ are different at different layers. 
The $T_i$ and $\beta_i$ maps at six different layers are Gaussian random numbers generated from six sets of pixel weighted average values over \planck\ estimate dust temperature and spectral index maps that follows auto and cross spectra of temperature and spectral index obtained from observed Planck dust map \citep{planck-XLVIII:2016}. Those fluctuations are further re-scaled to match the non-Gaussian distribution of \planck\ spectral maps and intensities at 353, 545, 857 GHz and 100 microns. We estimate this scaling method introduces an amount of 0.5\%  decorrelation at $\ell = 80$ between 217 and 353 GHz over 70 \% of the unmasked sky. 


\subsection{Instrumental Noise}
\label{ssec:Noise}
\label{sec:noisesims}

In this paper, we assume the instrumental noise to be Gaussian, white and uncorrelated across the frequency channels. \rev{We generate a random realization of a noise map} and project onto a \healpix\ map of \Nside\ = 256. The root mean square value of noise at each pixel $\sigma_{pix}^i$ is directly related to the sensitivity $s^i$ at respective channels as \citep{PhysRevD.52.4307}:
\begin{equation}
    \sigma_{pix}^i= \sqrt{\frac{(s^i)^2}{\Omega_{pix}}}.
\end{equation}
Here, $\Omega_{pix}$ is the area of each pixel of the map in $arcmin^2$. 

To minimize the level of noise contamination in the calculation of the CMB angular power spectrum, we simulate two \enquote{half-mission} maps. To simulate two uncorrelated sets of noise realizations, we use the sensitivity of $\sqrt{2}s^i$ at respective frequency channels. To obtain the corresponding full-mission noise realizations, we average over the two half-mission noise maps. Noise realizations of the full mission and two \enquote{half-mission} surveys are coadded with CMB and foreground maps to obtain the final \ECHO\ frequency maps. 

\section{Methodology}
\label{sec:methodology}
\begin{figure}
\includegraphics[width=\columnwidth]{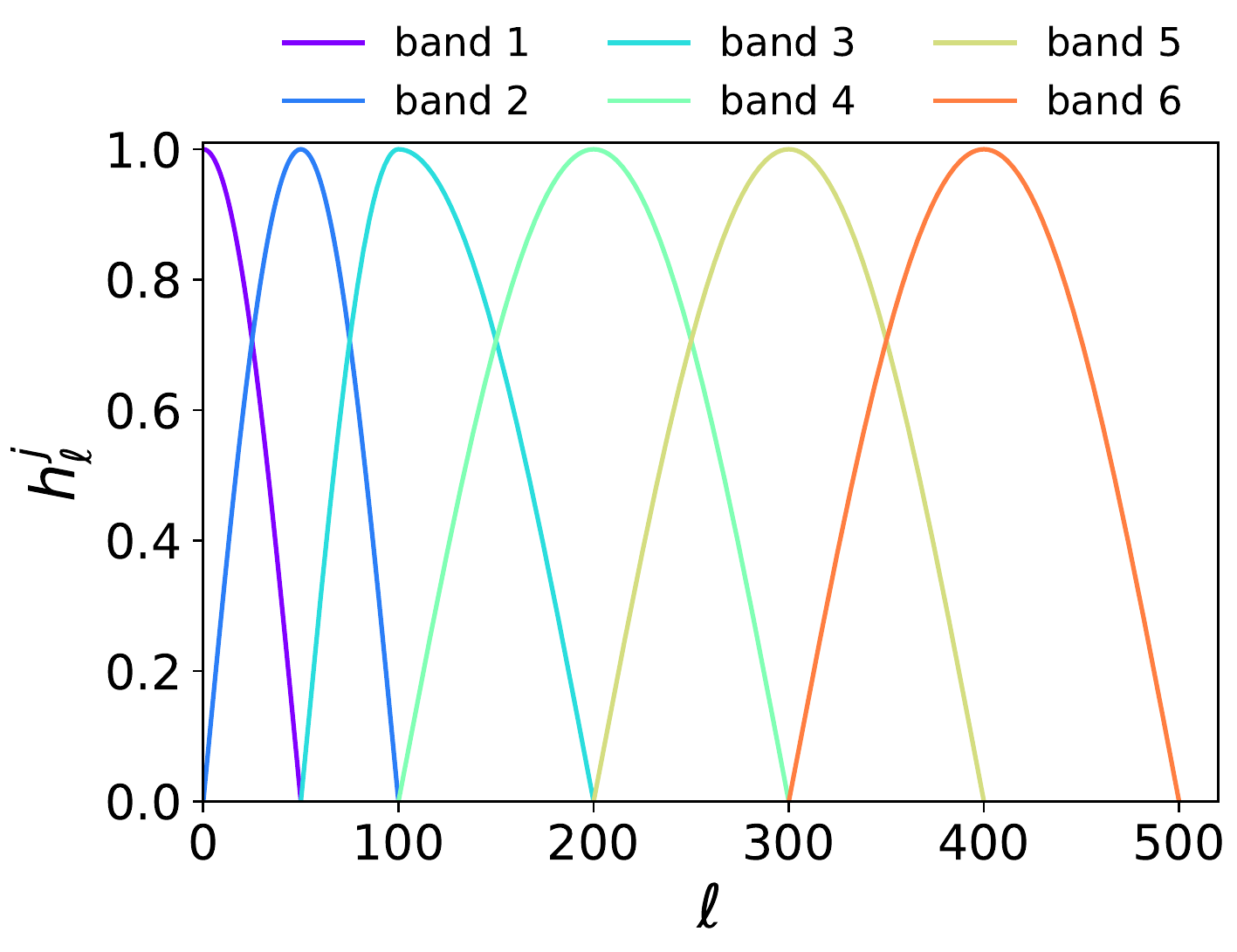}      
\caption{The cosine needlet filter function $h^{j}_{\ell}$ that is used to decompose the input sky maps into six needlet bands. The lower needlet bands carry the large-scale information and are crucial for foreground removal, while the higher needlet bands are vital for the minimization of instrumental noise.  }
\label{fig:NILC_filter}
\end{figure}

In this section, we describe the methodology used to forecast \ECHO's capability in measuring the value of $r$. 
To extract the CMB from the sky simulations, we use the blind component separation method, \NILC.\ The method is an implementation of the internal linear combination (ILC) on a needlet basis. Since the \NILC\ pipeline is applicable to only scalar fields, we first extract the $B$-mode of polarization from the simulations of $Q$ and $U$. After this, a cosine filter function $h_{\ell}^{j}$, where $j$ stands for the needlet scale/band, is used to decompose the $B$-mode maps at $\Nside=256$ into needlet coefficient maps at six needlet bands. The needlet bands and their corresponding \healpix\ resolution of the needlet coefficient maps are listed in Table~\ref{tab:needlet-bands}, and the filter function is depicted in Fig.~\ref{fig:NILC_filter}. In order to clean the foreground and noise, the multifrequency needlet coefficient maps are linearly combined through the \NILC\ weights. Since the needlet basis is localized in pixel space, it allows the \NILC\ weights to adapt for the local variation in foreground and noise as well as smooth variation in large and small harmonic scale to subtract the foreground and noise simultaneously.  Finally, the \NILC\ recovered CMB map is obtained through the synthesis of the cleaned needlet coefficient maps. The final \NILC\ maps will also have contributions from foreground and noise (residuals). \reff{Lower residual levels imply better performance of} \NILC.\ The \NILC\ pipeline allows us to estimate residual foreground and noise independently. This is achieved by propagating the \NILC\ weights to simulations of the foreground and noise. 
We implement the \NILC\ pipeline on the \enquote{full-mission}, as well as the two \,\enquote{half-mission} simulations.
%
\begin{table}
\caption{The needlet band index,j, and the corresponding \Nside\ at which the needlet coefficient maps are obtained from the input sky maps. The cosine filter function $h_{\ell}^{j}$ is used to decompose the sky maps into needlet bands. The function has compact support in the multipole range of $[\ell_{min},\ell_{max}]$ and has a maximum value of $h_{\ell}^{j}=1$ at $\ell_{peak}$.}  
\label{tab:needlet-bands} 
\begin{tabular}{c c c c c} 
\hline\hline \noalign{\vskip 2pt}
  Band index,j & $\ell_{min}$ & $\ell_{peak}$ & $\ell_{max}$ & \Nside \\ [1ex]
  \hline 
  1 & 0 & 0 & 50 & 32\\
  2 & 0 & 50 & 100 & 64 \\
  3 & 50 & 100 &
  200 & 128 \\
  4 & 100 & 200 & 300 & 128 \\
  5 & 200 & 300 & 400 & 256\\
  6 & 300 & 400 & 500 & 256\\  
  \hline
  \hline
\end{tabular} 
\end{table}

Next, we estimate the cross-angular power spectrum  over apodized mask from the two half-mission CMB $B$-mode maps through the pseudo-$C_{\ell}$ method ~\citep{Tegmark:1997}. The preparation and optimization of the sky mask are discussed in Sec.~\ref{ssec:sky-fraction}. We use the publicly available code \textit{Xpol}\footnote{\url{https://gitlab.in2p3.fr/tristram/Xpol}} \citep{10.1111/j.1365-2966.2005.08760.x} to obtain unbiased estimates of $\mathcal{D}_{\ell}$ and covariance matrices (denoted by $\Xi$) corrected for the effects of incomplete sky coverage. Here $\mathcal{D}_{\ell}$ depends on angular power spectrum as $\ell(\ell+1)C_{\ell}/2\pi$. Since the power spectra are derived from \enquote{half-mission}  maps with uncorrelated noise, any bias in its estimate will be due to the cosmic variance and the residual foreground. To calculate the covariance matrix, \textit{Xpol} uses an analytical approach, and the contributions from (residual) noise, foreground and cosmic variance are taken into account. To minimize the effect of the correlation 
along neighboring multipoles, we adopt a binning scheme with a bandwidth of $\Delta \ell=9$ in our calculations.  

Apart from this, we also calculate the binned angular power spectra from the masked residual foreground and noise maps denoted by $\hat{\mathcal{D}}_b^{\mathrm{FG}}$ and $\hat{\mathcal{D}}_b^{\mathrm{NS}}$ respectively using \textit{Xpol}. 

Finally, we estimate the tensor-to-scalar ratio using the $BB$ angular power spectrum-based likelihood method. The log of the likelihood can be expressed as;
\begin{equation}
\label{eq:likelihood}
\begin{aligned}
    -2\mathcal{L} =\, & constant +\\
     & \sum_{b,b'} \left( \hat{\mathcal{D}}_{b}^{data} - \mathcal{D}_{b}^{th} (r)\right ) \hat{\Xi}_{b,b'}^{-1}\left( \hat{\mathcal{D}}_{b'}^{data} - \mathcal{D}_{b'}^{th} (r)\right ).
    \end{aligned}
\end{equation}

Here, $\hat{\mathcal{D}}_{b}^{data}$ is the binned power spectra derived from the recovered CMB maps. $\hat{\Xi}_{b,b'}$ is the noise covariance matrix. 
The theoretical CMB $B$-mode power spectrum $\mathcal{D}_{b}^{BB,th}$, has contributions both from the lensing $B$-mode spectra and the tensor modes. It can be modeled as;
\begin{equation}
    \mathcal{D}_{b}^{th} (r) = \frac{r}{r_0} \mathcal{D}_{b}^{tensor} (r = r_0) +  A_{L}\mathcal{D}_{b}^{lensing}.
    \label{eq:r-model} 
\end{equation}
Here, $r_0$ is the reference tensor-to-scalar ratio and $A_{L}=1$, implying we have not used delensed maps. The tensor part of the $BB$-power \reff{spectrum} scales linearly with $r$. The angular power spectra, $\mathcal{D}_{b}^{lensing}$ and $\mathcal{D}_{b}^{tensor}$ are calculated from the \planck\ cosmological parameters of the $\Lambda$CDM-model \citep{planck-IV:2018} using \camb. The mean level of the lensing contribution $\mathcal{D}_{b}^{lensing}$ is subtracted at the power spectrum level, and the corresponding error bars are added to the noise covariance matrix $\hat{\Xi}_{b,b'}$ such that
\begin{equation}
    \Xi_{b,b'} = \hat{\Xi}_{b,b'} + \Xi_{b,b'}^{lensing}.
\end{equation}
The most probable value of the tensor-to-scalar ratio, $r_{\mathrm{mp}}$ can be obtained by maximizing the likelihood function in Eq.~\ref{eq:likelihood} with respect to $r$ as follows,
\begin{equation}
    r_{\mathrm{mp}} = r_{0}\times \frac{{\sum}_{bb'} \left( \hat{\mathcal{D}}_b^{data}- \mathcal{D}_b^{lensing}\right)\Xi_{b,b'}^{-1}\mathcal{D}_{b'}^{tensor}}{{\sum}_{bb'}\mathcal{D}_b^{tensor}\,\Xi_{b,b'}^{-1}\,\mathcal{D}_{b'}^{tensor}}.
    \label{eq:r-mp} 
\end{equation}
In our analysis, we set $r_{0}=1$ and obtain $\hat{\mathcal {D}}_{b}^{data}$ by calculating the cross-power spectra from the \NILC\ recovered \enquote{half-mission} 1 and 2 maps. The estimation of $r_{\mathrm{mp}}$ will be biased due to cosmic variance and the presence of residual foreground in the recovered maps. Since we are taking the cross spectra between the two \enquote{half-mission} maps, we are able to avoid the effect of residual noise in the recovered $r_{\mathrm{mp}}$ values. The uncertainty on $r_{\mathrm{mp}}$ can be estimated from the second-order derivative of the likelihood function and depends on the covariance matrix $\Xi$. This is given by;
\begin{equation}
    \sigma_{r} = r_{0}\times \left( \sum_{bb'}\mathcal{D}_b^{tensor}\,\Xi_{b,b'}^{-1}\,\mathcal{D}_{b'}^{tensor}\right)^{-1/2}. 
    \label{eq:r-sig} 
\end{equation}
Since the pseudo-$C_{\ell}$ method is not accurate at lower multipoles~\citep{10.1111/j.1365-2966.2005.08760.x,10.1111/j.1365-2966.2004.07530.x}, we consider multipoles $\ell\ge30$ for the $r_{\mathrm{mp}}$ and $\sigma_{r}$ estimation. This strategy also helps us in mitigating systematic errors in parameter estimation due to the non-Gaussian likelihood of the power spectra at lower multipoles~\citep{10.1111/j.1365-2966.2006.10910.x}.

%

\section{Results}
\label{sec:results}
The goal of our work is to understand how \ECHO's sensitivity  s the detection of $r$  changes with the inclusion of its high-frequency bands. For this, we have applied the \NILC\ pipeline to extract the CMB $B$-mode from the sky simulations for various configurations of \ECHO\ frequency bands. In this section, we present the  results obtained from the \NILC\ recovered CMB and residual maps. 
\subsection{Figures of merit}
\label{ssec:FOM}

To assess the performance of  \NILC\ pipeline for various instrument configurations of \ECHO, we examine the level of foreground and noise residuals for three different sets of sky simulations. 
We define the average of the binned angular power spectra of the noise and foreground residuals as the figures of merit: $\bar{\mathcal{D}}^{\mathrm{NS}}$(FoM-1) and $\bar{\mathcal{D}}^{\mathrm{FG}}$(FoM-2);

\begin{align}
\bar{\mathcal{D}}^{\mathrm{NS}} \equiv \frac{1}{N_{bin}} \sum_{b=b_{min}} ^{b_{max}}\ \hat{\mathcal{D}}_{b}^{\mathrm{NS}} \ , 
\end{align}
\begin{align}
\bar{\mathcal{D}}^{\mathrm{FG}} \equiv \frac{1}{N_{bin}} \sum_{b=b_{min}} ^{b_{max}}\ \hat{\mathcal{D}}_{b}^{\mathrm{FG}}.
\end{align}
To estimate these figures of merit, we have considered the multipole range of $\ell=[30,200]$.

\subsection{Sky fraction}
\label{ssec:sky-fraction}

\begin{figure}
    \includegraphics[width=\columnwidth]{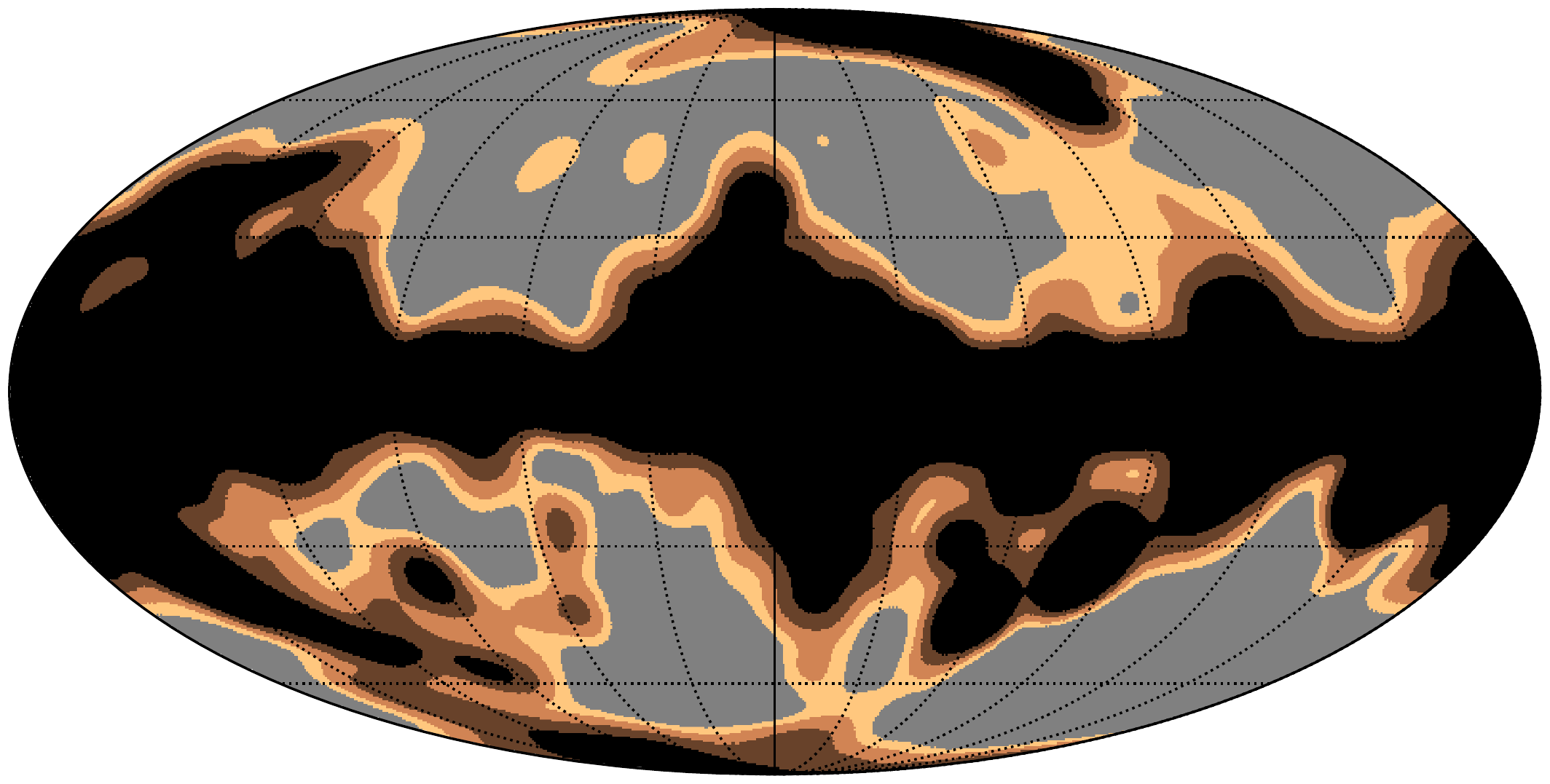}\par
       
    \caption{The binary masks used for estimating $B$-mode power spectrum. \reff{The darkest brown shaded region represents the portion of the sky removed from the analysis to retain the sky fraction, $SF=60\%$. The three consecutive fainter shades of brown color represent the regions incrementally masked out for retaining 50\%, 40\%, and 30\% of the sky.} These masks are generated from smoothed residual foreground maps derived by applying the \NILC\ method to simulations of the sky for the $20$ \ECHO\ bands of the M2 sky model.}
    
    \label{fig:nilc_mask}
\end{figure}
\begin{figure*}[!htbp]
\includegraphics[width=\linewidth, height=0.3\textwidth, keepaspectratio]{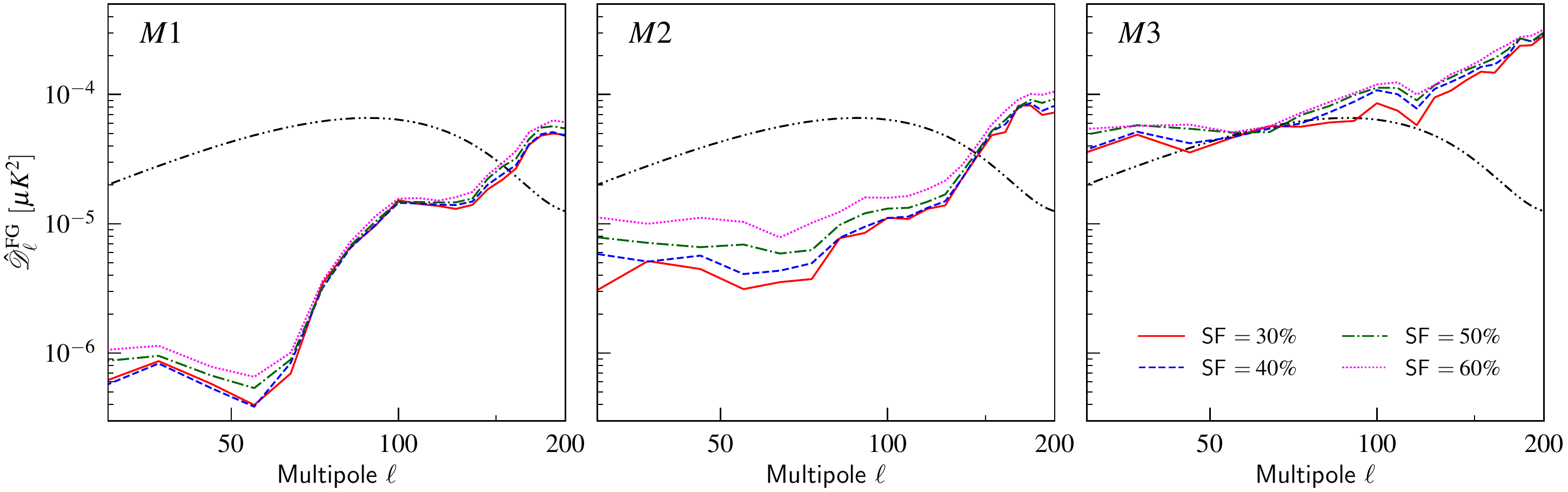}
\caption{The residual foreground spectra for sky fraction of $30\%-60\%$. The power of the foreground residuals decreases as we use more conservative masks. From \textit{left} to \textit{right} we depict the residual levels for $M1$, $M2$, and $M3$, respectively. The black dash-dot-dot lines show the tensor mode of $BB$ theoretical power spectrum for $r=10^{-3}$.}
\label{fig: SF-residual-fg}
\end{figure*}

\begin{figure*}[!htbp]
\includegraphics[width=\linewidth, height=0.3\textwidth, keepaspectratio]{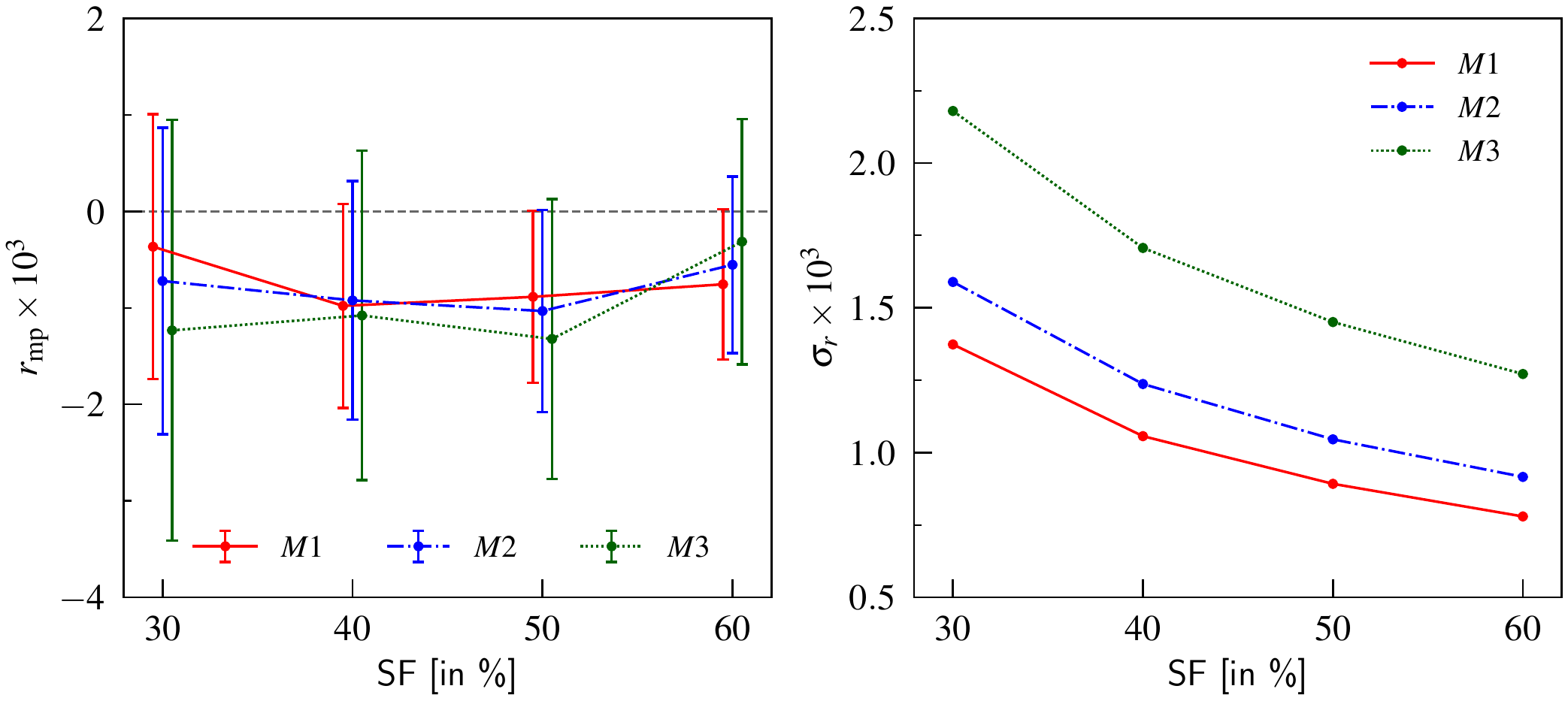}
\caption{\textit{Left panel}: The recovered $r_{\mathrm{mp}}$ values and their uncertainty with respect to the sky fraction. The bias in $r_{\mathrm{mp}}$ has similar values for the sky coverage between 30-60\%. The data points are shifted in the X-axis for clarity. \textit{Right panel:} The increase in the uncertainty, $\sigma_{r}$ as the sky fraction is reduced.}
\label{fig: SF-r-sigma}
\end{figure*}

To reduce the impact of foreground residuals on the estimation of $r$, we use a set of conservative masks.  These masks are produced from the foreground residual maps obtained by implementing \NILC\ on the $M2$ sky simulations.
The variance of the foreground residual maps is then smoothed by a Gaussian beam of FWHM = $9^{\circ}$. The final masks of desired \textit{Sky Fraction} (SF) are obtained using a suitable threshold on these smoothed maps. Fig.~\ref{fig:nilc_mask} shows binary masks with the sky coverage of $30\%,\,40\%,\,50\%$ and $60\%$.
Prior to applying these masks on the recovered CMB maps to estimate angular power spectra, we apodize the masks with Gaussian kernel of FWHM = $2^{\circ}$.


To optimize the mask for our data analysis, we closely examine the foreground residuals, $r_{\mathrm{mp}}$ values, and corresponding uncertainties over different masks.  In Fig.~\ref{fig: SF-residual-fg}, we present the residual foreground power spectra over four different masks. We find for all three sky models; the residual foreground power spectra foreseeably decrease monotonically as we reduce the sky coverage. In Fig.~\ref{fig: SF-r-sigma} we present estimated $r_{\mathrm{mp}}$ (\textit{left panel}) and corresponding $\sigma_r$ (\textit{right panel}) over the same set of masks. The recovered $r_{\mathrm{mp}}$ values for different sky models are close to each other for sky fraction of $30-60\%$ and lie within the $1\sigma$ limit. At the same time, the uncertainties increase as we reduce the sky fraction. Therefore, we choose 60\% of sky coverage which gives the minimum value of $\sigma_r$, henceforth in our work. 

\subsection{Results from the proposed instrument configuration}
\label{ssec:full echo}
\begin{figure*}[!htbp]
\includegraphics[width=\linewidth, height=0.4\textwidth, keepaspectratio]{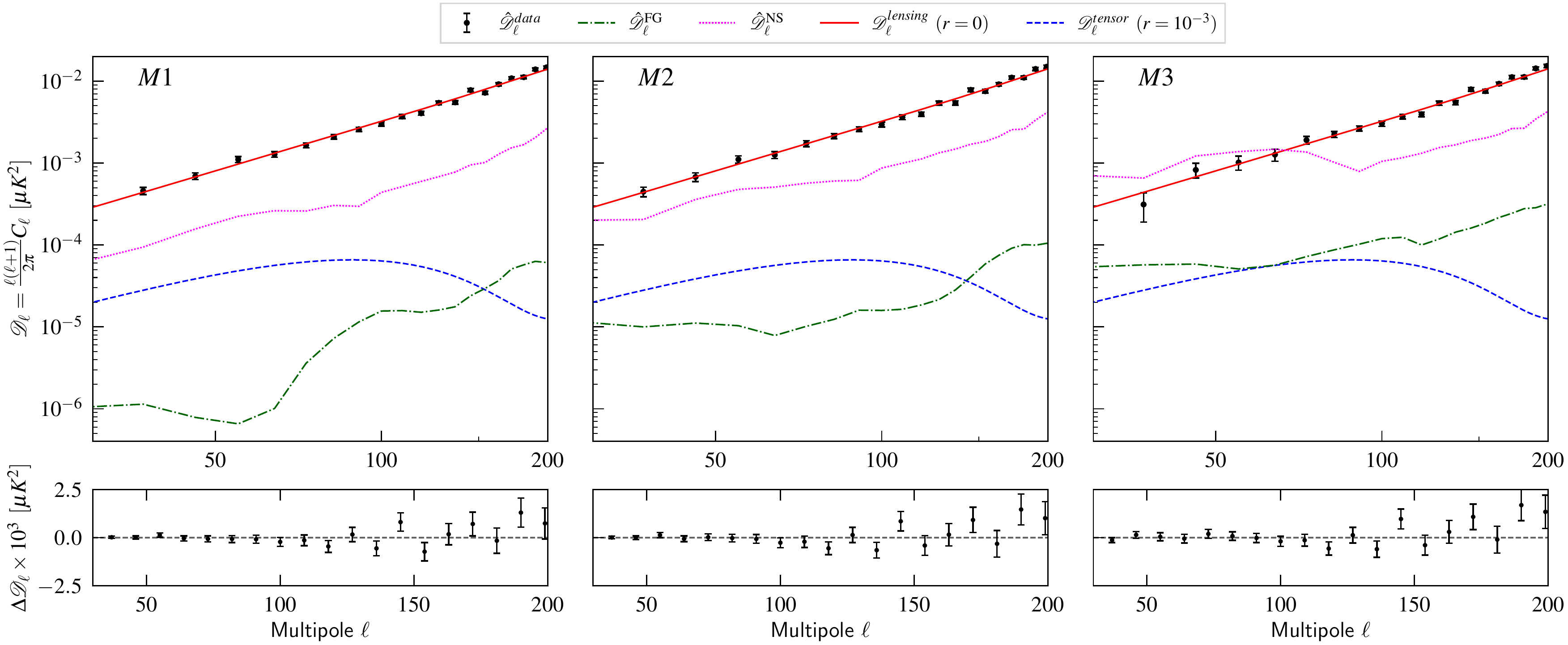}
\caption{\textit{Left to right:} the plot of the \NILC\ recovered CMB, residual foregrounds, and residual noise $BB$ angular power spectra $\mathcal{D}_{\ell} = \frac{\ell(\ell+1)}{2\pi}C_{\ell}$ for three set of sky simulations. \textit{Upper panel:} the black dots depict the cross angular power spectra from the \enquote{half-mission} \NILC\ recovered CMB maps, the green lines show the residual foregrounds and the magenta lines are for the residual noise levels. The red lines represent the theoretical CMB lensed power spectra for $r_{in}=0$, and the blue lines represent the tensor theoretical spectra for $r=10^{-3}$, which is the target sensitivity of \ECHO. \textit{Lower panel:} plot of recovered CMB tensor power spectrum, $\Delta \mathcal{D}_{\ell} = \hat{\mathcal{D}}_b^{data}- \mathcal{D}_b^{lensing}$, obtained by subtracting the theoretical lensed spectrum from \NILC\ recovered power spectra. The error bars corresponding to the lensed power spectrum are added to the \NILC\ recovered error bars. It is consistent with $r_{in}=0$ within $2 \sigma$ error bars.} 
\label{fig:fiducial}
\end{figure*}


In order to demonstrate the model complexity considered here, we first assess the \NILC\ results for the \ECHO's\ fiducial configuration with all $20$ frequency bands listed in  Table~\ref{table:r.m.s.}. Fig.~\ref{fig:fiducial} displays the recovered CMB $BB$ power spectrum for the three sky models. We find for all three sky models, the recovered CMB $BB$ power spectrum for $r_{in}$ = 0 is consistent with the theoretical lensed power spectrum within $1\sigma$ error bars. Furthermore, for the $M1$ and $M2$ models, the level of the residual foreground power spectrum is well below the theoretical tensor power spectrum for $r=10^{-3}$ in the multipole range of $30\leq \ell \leq 150$ and near \reff{the recombination peak}. However, the $M1$ model exhibits one order of magnitude less foreground residual compared to the $M2$ model. 
On the contrary, for the $M3$ model, the foreground residual power spectrum is much higher than in the previous two models. For all three models, foreground residual levels are much lower than residual noise levels. We also find that for the $M1$ model, the amplitude of   foregrounds and noise residuals is subdominant as compared to the lensed $BB$ power spectrum and lowest compared to the other two models. For the $M2$ model \reff{residual noise} is comparable to lensing power.  On the other hand, for the $M3$ model, \reff{the residual noise is higher than the lensing power at low multipoles.}

To quantify these results, in Table~\ref{table:fiducialFOM}, we have listed the values of FoM-1 and FoM-2 for all three sky models. We confirm that the residuals are the lowest for the $M1$ model. Compared to the $M1$ model, for the $M2$ model, both FoMs increase by ~$50$\% while the same for the $M3$ model increase by more than $100$\%. As a response, we see a systematic increase in the $\sigma_{r}$ values from $M1$ to $M3$ model that is shown in Table~\ref{table:fiducialFOM}. The recovered $r_{\mathrm{mp}}$ for all models is consistent with null detection within respective $1\sigma$ uncertainty. These results clearly show that \NILC\ is very efficient in cleaning the complexities introduced to the physical properties of dust in the $D1$ model, whereas subtraction of the multi-layer dust in the $D3$ model is more challenging. 
\begin{center}
\begin{table*}[!htbp]
\caption{The recovered tensor-to-scalar ratio $r_{\mathrm{mp}}$, $\sigma_{r}$ uncertainty, and the two FoMs based on residual noise and foreground for the fiducial \ECHO\ configuration.}
\label{table:fiducialFOM}
\begin{tabular} { c  c c c c}
\hline\hline \noalign{\vskip 2pt}
 Sky Model &\hspace{12ex}  $r_{\mathrm{mp}} \times 10^{3}$ &\hspace{12ex}  $\sigma_{r} \times 10^{3}$ &\hspace{12ex}  \thead{$\bar{\mathcal{D}}^{\mathrm{NS}} \times 10^{3}$\\$[\mu K^2]$} &\hspace{12ex}  \thead{$\bar{\mathcal{D}}^{\mathrm{FG}} \times 10^{4}$\\$[\mu K^2]$}\\
\noalign{\vskip 2pt}\hline \noalign{\vskip 2pt}
\rule[-0ex]{0pt}{0ex}$M1$ &\hspace{12ex} -0.76 &\hspace{12ex} 0.78 &\hspace{12ex} 0.7 &\hspace{12ex} 0.2\\
\rule[-0ex]{0pt}{0ex}$M2$ &\hspace{12ex} -0.55 &\hspace{12ex} 0.92 &\hspace{12ex} 1.2 &\hspace{12ex} 0.3\\
\rule[-0ex]{0pt}{0ex}$M3$ &\hspace{12ex} -0.31 &\hspace{12ex} 1.27 &\hspace{12ex} 1.6 &\hspace{12ex} 1.3\\
\hline\hline
\end{tabular}
\end{table*}
\end{center}

\subsection{Inclusion of dust dominated bands}
\begin{figure*}[!htbp]
\includegraphics[width=\linewidth, height=0.7\textwidth, keepaspectratio]{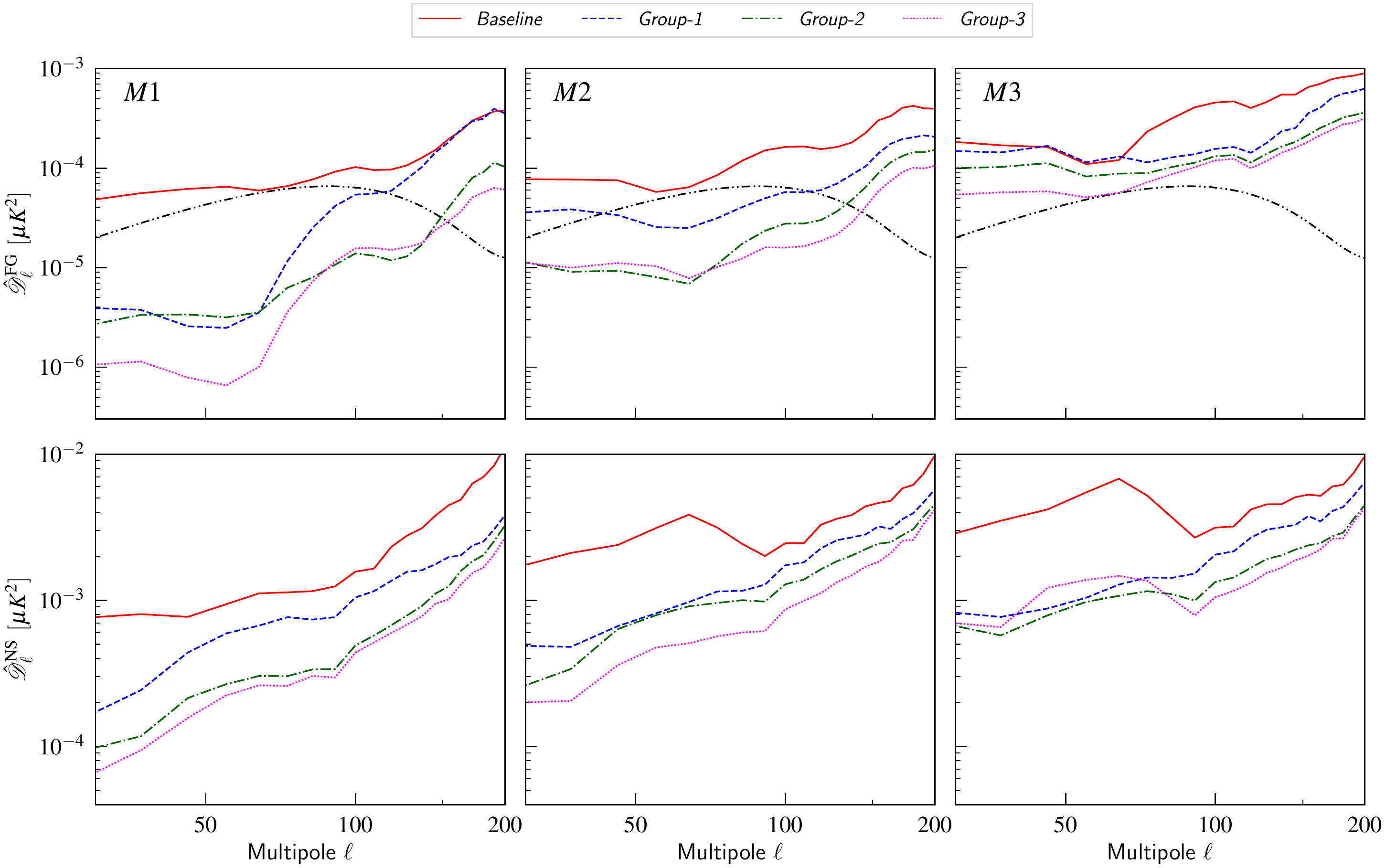}
\caption{\label{fig:residuals}The $BB$ angular power spectrum for the residual foreground (\textit{upper panel}) and residual noise (\textit{lower panel}) for the four groups of \ECHO\ bands. \reff{The \textit{pink} lines correspond to the baseline configuration, which includes 11 bands ranging from $28-190$\GHz\ (covering the least frequency range). The \textit{green} line illustrates the residuals for the configuration with 14 frequency bands spanning from $28$ \GHz\ to $340$ \GHz.\ The \textit{blue} line represents the residuals for the configuration with 18 frequency bands covering the range of $28-600$ \GHz\ and the \textit{red} line represents the residuals for the configuration in \textit{Group-3}, which consists of 20 frequency bands in the $28-850$ \GHz\ range. }The dashed-dot-dot lines in the upper panel depict the power of theoretical tensor $BB$ spectra for $r=10^{-3}$.}
\end{figure*}

\begin{table*}[!htbp]
\caption{The FoM-1, $\bar{\mathcal{D}}^{\mathrm{NS}}$ values obtained from the residual noise spectra $\hat{\mathcal{D}}_{b}^{\mathrm{NS}}$ and the corresponding configurations of \ECHO\ bands. We have also calculated their percentage change with respect to the FoM-1 obtained from configuration-1.}
\label{table:FoM1}
\begin{tabular} { c  c  c c c c c c}
\hline\hline \noalign{\vskip 2pt}
\rule[-1.5ex]{0pt}{1.ex} & \rule[-1.5ex]{0pt}{1.ex} &  \multicolumn{2}{c}{$M1$} &  \multicolumn{2}{c}{$M2$} &  \multicolumn{2}{c}{$M3$}\\
 \thead{Configuration} &\hspace{0.5ex}  \thead{Frequency \\range (GHz)} &\hspace{0.5ex}  \thead{$\bar{\mathcal{D}}^{\mathrm{NS}} \times 10^{3}$\\$[\mu K^2]$} &\hspace{0.5ex}  $\%$ Change &\hspace{0.5ex}  \thead{$\bar{\mathcal{D}}^{\mathrm{NS}} \times 10^{3}$\\$[\mu K^2]$} &\hspace{0.5ex}  $\%$ Change &\hspace{0.5ex}  \thead{$\bar{\mathcal{D}}^{\mathrm{NS}} \times 10^{3}$\\$[\mu K^2]$} &\hspace{0.5ex}  $\%$ Change\\
\noalign{\vskip 2pt}\hline \noalign{\vskip 2pt}
\rule[-0ex]{0pt}{0ex}\textit{Baseline} &\hspace{0.5ex} \rule[-0ex]{0pt}{0ex}28-190 &\hspace{0.5ex} 2.8 &\hspace{0.5ex} 0 &\hspace{0.5ex} 3.7 &\hspace{0.5ex} 0 &\hspace{0.5ex} 4.7 &\hspace{0.5ex} 0\\
\rule[-0ex]{0pt}{0ex}\textit{Group-1} &\hspace{0.5ex} \rule[-0ex]{0pt}{0ex}28-340 &\hspace{0.5ex} 1.3 &\hspace{0.5ex} 53 &\hspace{0.5ex} 2.1 &\hspace{0.5ex} 43 &\hspace{0.5ex} 2.4 &\hspace{0.5ex} 48\\
\rule[-0ex]{0pt}{0ex}\textit{Group-2} &\hspace{0.5ex} \rule[-0ex]{0pt}{0ex}28-600 &\hspace{0.5ex} 0.8 &\hspace{0.5ex} 71 &\hspace{0.5ex} 1.6 &\hspace{0.5ex} 56 &\hspace{0.5ex} 1.7 &\hspace{0.5ex} 63\\
\rule[-0ex]{0pt}{0ex}\textit{Group-3} &\hspace{0.5ex} \rule[-0ex]{0pt}{0ex}28-850 &\hspace{0.5ex} 0.7 &\hspace{0.5ex} 75 &\hspace{0.5ex} 1.2 &\hspace{0.5ex} 67 &\hspace{0.5ex} 1.6 &\hspace{0.5ex} 65\\
\hline\hline
\end{tabular}
\end{table*}

\begin{table*}[!htbp]
\caption{The FoM-2, $\bar{\mathcal{D}}^{\mathrm{FG}}$ values obtained and its percentage change with respect to FoM-2 obtained from configuration-1.}
\label{table:FoM2}
\begin{tabular} { c  c  c c c c c c}
\hline\hline \noalign{\vskip 2pt}
\rule[-1.5ex]{0pt}{1.ex} & \rule[-1.5ex]{0pt}{1.ex} &  \multicolumn{2}{c}{$M1$} &  \multicolumn{2}{c}{$M2$} &  \multicolumn{2}{c}{$M3$}\\
 \thead{Configuration} &\hspace{0.5ex}  \thead{Frequency \\range (GHz)} &\hspace{0.5ex}  \thead{$\bar{\mathcal{D}}^{\mathrm{FG}} \times 10^{4}$\\$[\mu K^2]$} &\hspace{0.5ex}  $\%$ Change &\hspace{0.5ex}  \thead{$\bar{\mathcal{D}}^{\mathrm{FG}} \times 10^{4}$\\$[\mu K^2]$} &\hspace{0.5ex}  $\%$ Change &\hspace{0.5ex}  \thead{$\bar{\mathcal{D}}^{\mathrm{FG}} \times 10^{4}$\\$[\mu K^2]$} &\hspace{0.5ex}  $\%$ Change\\
\noalign{\vskip 2pt}\hline \noalign{\vskip 2pt}
\rule[-0ex]{0pt}{0ex}\textit{Baseline} &\hspace{0.5ex} \rule[-0ex]{0pt}{0ex}28-190 &\hspace{0.5ex} 1.4 &\hspace{0.5ex} 0 &\hspace{0.5ex} 1.9 &\hspace{0.5ex} 0 &\hspace{0.5ex} 4.4 &\hspace{0.5ex} 0\\
\rule[-0ex]{0pt}{0ex}\textit{Group-1} &\hspace{0.5ex} \rule[-0ex]{0pt}{0ex}28-340 &\hspace{0.5ex} 1.1 &\hspace{0.5ex} 21 &\hspace{0.5ex} 0.9 &\hspace{0.5ex} 52 &\hspace{0.5ex} 2.4 &\hspace{0.5ex} 45\\
\rule[-0ex]{0pt}{0ex}\textit{Group-2} &\hspace{0.5ex} \rule[-0ex]{0pt}{0ex}28-600 &\hspace{0.5ex} 0.3 &\hspace{0.5ex} 78 &\hspace{0.5ex} 0.5 &\hspace{0.5ex} 73 &\hspace{0.5ex} 1.6 &\hspace{0.5ex} 63\\
\rule[-0ex]{0pt}{0ex}\textit{Group-3} &\hspace{0.5ex} \rule[-0ex]{0pt}{0ex}28-850 &\hspace{0.5ex} 0.2 &\hspace{0.5ex} 85 &\hspace{0.5ex} 0.3 &\hspace{0.5ex} 84 &\hspace{0.5ex} 1.3 &\hspace{0.5ex} 70\\
\hline\hline
\end{tabular}
\end{table*}

\begin{table*}[!htbp]
\caption{The inferred $r_{\mathrm{mp}}$ for input $r$ = 0 and 1$\sigma$ uncertainty for four selected configurations of frequency bands.}
\label{table:opt-r-sigma}
\begin{tabular} { c  c  c c c c c c}
\hline\hline \noalign{\vskip 2pt}
\rule[-1.5ex]{0pt}{1.ex} & \rule[-1.5ex]{0pt}{1.ex} &  \multicolumn{2}{c}{$M1$} &  \multicolumn{2}{c}{$M2$} &  \multicolumn{2}{c}{$M3$}\\
 \thead{Configuration} &\hspace{5ex}  \thead{Frequency \\range (GHz)} &\hspace{5ex}  $r_{\mathrm{mp}} \times 10^{3}$ &\hspace{5ex}  $\sigma_{r} \times 10^{3}$ &\hspace{5ex}  $r_{\mathrm{mp}} \times 10^{3}$ &\hspace{5ex}  $\sigma_{r} \times 10^{3}$ &\hspace{5ex}  $r_{\mathrm{mp}} \times 10^{3}$ &\hspace{5ex}  $\sigma_{r} \times 10^{3}$\\
\noalign{\vskip 2pt}\hline \noalign{\vskip 2pt}
\rule[-0ex]{0pt}{0ex}\textit{Baseline} &\hspace{5ex} \rule[-0ex]{0pt}{0ex}28-190 &\hspace{5ex} -0.54 &\hspace{5ex} 1.24 &\hspace{5ex} 1.04 &\hspace{5ex} 1.91 &\hspace{5ex} 4.59 &\hspace{5ex} 2.42\\
\rule[-0ex]{0pt}{0ex}\textit{Group-1} &\hspace{5ex} \rule[-0ex]{0pt}{0ex}28-340 &\hspace{5ex} -0.39 &\hspace{5ex} 0.95 &\hspace{5ex} -0.73 &\hspace{5ex} 1.14 &\hspace{5ex} 0.98 &\hspace{5ex} 1.33\\
\rule[-0ex]{0pt}{0ex}\textit{Group-2} &\hspace{5ex} \rule[-0ex]{0pt}{0ex}28-600 &\hspace{5ex} -0.49 &\hspace{5ex} 0.81 &\hspace{5ex} -0.39 &\hspace{5ex} 1.05 &\hspace{5ex} 0.58 &\hspace{5ex} 1.21\\
\rule[-0ex]{0pt}{0ex}\textit{Group-3} &\hspace{5ex} \rule[-0ex]{0pt}{0ex}28-850 &\hspace{5ex} -0.76 &\hspace{5ex} 0.78 &\hspace{5ex} -0.55 &\hspace{5ex} 0.92 &\hspace{5ex} -0.31 &\hspace{5ex} 1.27\\
\hline\hline
\end{tabular}
\end{table*}

\reff{The thermal dust is the dominating foreground emission at frequencies above 70 GHz and increases rapidly with frequency. 
This makes the high-frequency observations of the dust foregrounds very important for their subtraction in the component separation exercise. In fact, nine channels in the range of $220-850$ \GHz\ have been included in the design of \ECHO\ mission for this purpose. However, the downside of this approach is that the inclusion of high-frequency channels will make the size of the detector plane bigger, which ultimately will enhance the cost of the mission significantly. It is important to analyze the trade-off between the cost of the mission and an optimum number of channels to be placed.}

In this section, we explore the utility of high-frequency coverage for increasing the \rev{detection significance of the $B$-modes of the CMB}. In order to do this, we initially examine the most economical setup, which involves utilizing $11$ frequency bands ranging from $28$  to $190$ \GHz\ as our baseline configuration. We note that selecting a baseline of $28-220$\GHz\ would fulfil this requirement as well, and either option does not alter the primary findings of our work. 
Subsequently, we explore the possibility of improvements in that direction by extending the frequency range to higher frequencies. We assess $3$ sets of frequency bands with an increasing number of dichroic pixels. The first group, \textit{Group-1}, comprises $14$ frequency bands ranging from $28-340$ \GHz.\ The next group (\textit{Group-2} in our case) includes $18$ bands ranging from $28$ \GHz\ to $600$ \GHz.\ The final group (\textit{Group-3}) consists of $20$ \ECHO\ bands for the range of $28-850$ \GHz.
 

In Fig.~\ref{fig:residuals}, we demonstrate the level of residual foreground and noise for the groups of frequency bands mentioned above. The results clearly show that for all the dust models mentioned in section ~\ref{sec:thermaldust}, the level of residual foreground emissions and noise decreases with the increase of dust-dominated high-frequency bands. This is expected because \NILC\ uses the extra information gained upon adding the dust-dominated channels to subtract the dust component. 
However, the decline in the level of residual spectra is not uniform across the multipole range with addition of the higher frequency bands. To understand its dependency on the frequency range chosen more deeply, we analyze the trend found in the FoMs defined in Sec.~\ref{ssec:FOM}. In Table~\ref{table:FoM1} and Table~\ref{table:FoM2} we summarize the FoM-1 and FoM-2 values, respectively, for different groups and their percentage change with respect to the  baseline configuration. 

\textbf{\textit{Baseline configuration:}} \reff{First, we note that the level of residuals for the baseline configuration is the highest regardless of the dust model considered. This is expected because the number of channels, the dust-dominated channels, in particular, is minimum. 
We have listed these inferred $r_{\mathrm{mp}}$s and corresponding $\sigma_r$ values in Table~\ref{table:opt-r-sigma}. 
The measurement of $r$ \rev{for the $M1$ and $M2$ models} is consistent with null detection within the $1\sigma$ error bar. However, for the $M3$ model, $r_{\mathrm{mp}}$ bias is much higher as compared to the other two models. This is because of the frequency-frequency decorrelation present in the dust model used in $M3$.} 

{\textbf{\textit{Group-1}}}: We obtain a significant improvement in the FoM values \rev{upon addition of the $220$, $275$ and $340$ \GHz\ bands}. \rev{For the $M1$, $M2$, and $M3$ models}, we find that compared to the baseline configuration, FoM-1 decreases by a total of $53\%$, $43\%$ and $48\%$, respectively, when these three bands are included in the analysis. The FoM-2 for the $M1$ model  decreases by a total of $24\%$, while for $M2$ and $M3$, FoM-2 decreases by $55\%$ and $45\%$ respectively.
\reff{The sensitivity $\sigma_{r}$ in the measurement of the tensor-to-scalar ratio  \reff{improves as the level of} residuals decrease. For the $M1$ model it drops from $1.24\times 10^{-3}$ to $0.95 \times 10^{-3}$ implying an improvement of $\sim30$\%.  For $M2$ there is an improvement of $\sim40$\% for the same. 
Similarly, for $M3$, we note that the sensitivity improves by $\sim40\%$ as $\sigma_r$ decreases from $2.42\times 10^{-3}$ to $1.33\times 10^{-3}$. 
Furthermore, the bias in $r_{\mathrm{mp}}$ reduces considerably for the $M3$ models 
as compared to the same seen for the baseline configuration.} %


\textbf{\textit{Group-2:}} 
 \rev{We observe the additional changes in the \NILC\ results when} the next two sets of dichroic pixels, which are sensitive to $390$, $450$, $520$ and $600$ \GHz,\  are added. As listed in Table~\ref{table:FoM1}, we find that FoM-1 decreases by $17\%$, $13\%$ and $15\%$ for $M1$, $M2$ and $M3$ respectively. Consequently, we notice some improvements in the sensitivity as $\sigma_{r}$ that is improved by $\sim15\%$. For the frequency range of $28-520$\GHz,\ $\sigma_{r}=0.81\times10^{-3}$, $0.92\times10^{-3}$ and $1.28\times 10^{-3}$ for the $M1$, $M2$ and $M3$ models respectively. These values are pretty close to the uncertainty values obtained when observations \rev{until} maximum frequency coverage up to $850$ \GHz\ bands are used. Thus most of the improvements seen in $\sigma_{r}$ are already achieved when the observations \rev{until} $520$ \GHz\ are considered. 
In addition to this, FoM-2 is also reduced by $\sim57\%$ for the $M1$ model and about $20\%$ for the $M2$ and $M3$ models. The bias in the measurement of $r_{\mathrm{mp}}$ for all these cases is well  within $1\sigma$.

\textbf{\textit{Group-3:}}
\reff{Finally,  we assess the improvement of results with the addition of the last set of dichroic pixels, which are placed at central frequencies of 700 and 850 GHz. After the inclusion of these two final frequency bands, the level of noise residual reduces marginally. We find that the FoM-1 drops by an additional $5\%$, $10\%$ and $2\%$ for $M1$, $M2$ and $M3$ respectively. FoM-2 also falls off by about $5-8\%$. 
Similar to FoM-1, we find only marginal improvement in the $\sigma_r$ values. \rev{When increasing the frequency range} of observation from $600$\GHz\ to $850$\GHz\, $\sigma_r$ drops by an additional $3\%$ and $7\% $ \rev{for the $M1$ and $M2$ models, respectively}. For the $M3$ model, we do not see any improvement in sensitivity. We find nominal changes in the bias on $r_{\mathrm{mp}}$ 
for all three models, which are  due to both the change in the noise covariance matrix and the inherent statistical fluctuation associated with considering only one realization of the CMB map. However, the bias is consistent with null detection within $\pm 1\sigma_r$.} 

\section{Conclusions}
\label{sec:Conclusions}
Detecting the CMB $B$-mode is one of the primary goals of cosmology, and many space and ground-based experiments have been designed geared toward this challenge. \ECHO\ is a proposal for a fourth-generation space-based experiment, which will observe the microwave sky through 20 frequency channels at a very high resolution.

Removal of the dominant foreground from the observed signal is one of the major challenges in detecting $B$-mode. It is widely believed that increasing the frequency range at which the foreground is observed will directly improve the instrument's sensitivity toward $r$ recovery. This is the reason why future satellite experiments such as \ECHO\ and  \pico\ will have frequency bands up to $800-900$ GHz. In this work, we have tested this hypothesis in the context of thermal dust removal. We have used sky simulations comprising the CMB component with an input tensor-to-scalar ratio of $r_{in}=0$, synchrotron emission with power-law scaling, and thermal dust models with various forms of complexities. To understand the role of the dust-dominated frequency channels in the range of $220-850$ GHz, we observe how the performance of \NILC\ improves upon adding each dust monitor. We find the level of residuals systematically declines as we \reff{increase the frequency coverage by adding} the high-frequency bands, and as a response, the \reff{statistical uncertainty }also reduces in most cases. However, residual levels do not decrease uniformly \reff{with the addition of each high-frequency band and the resulting improvement seen in $\sigma_{r}$ can vary from $\sim30\%$ to just $2\%$.
We have found that for the $M1$ model, the residual noise level, which dictates the level of uncertainty, is reduced by a total of $75\%$ when we increase the frequency range from $28-190$\GHz\ to $28-850$\GHz.\ 
However, most of the contribution to this result has come from the first fourteen bands within the $28-340$ \GHz\ range. A similar pattern is also observed for the $M2$ and $M3$ models. Of the total $\sim 66\%$ of improvement seen in removing noise residuals, the majority (about $45\%$) is achieved by adding the first four dust-dominated bands in the $28-340$\GHz\ range. We also observe significant improvement in $r$ bias for the case of the sky model with decorrelated \GINES\ component (included in the $M3$ model). The addition of the subsequent four bands, spanning from $390-600$\GHz\, results in a further reduction of $15\%$ in the level of noise residuals for all the models. The uncertainty for the $M1$ model improves by an additional $20\%$. For the $M2$ and $M3$ models, the improvement in sensitivity is very small ($\sim 5\%$). Again the level of residuals decreases systematically by a small amount as we add the next two \ECHO\ bands in the range of 700-850 GHz. And consequently, we observe only marginal ($M1$ and $M2$ models) to no additional improvement in the uncertainty values, $\sigma_{r}$. }

\reff{Therefore, we conclude that \rev{the removal of the dust $B$-modes is sensitive to the choice of the frequency range.} The addition of dust-dominated channels at the high-frequency range does not necessarily translate into a proportional improvement in the detection sensitivity of $r$. \rev{The improvement seen as we add the highest frequency bands at} 700 and 850 \GHz\ is insignificant. \rev{However, to fully establish the redundancy of the 700 and 850 \GHz\ bands, we need to conduct additional analysis by maintaining consistency in the number of detectors and their corresponding sensitivity across each band. We intend to do this in the future.} These results presented here can be used to improve the \ECHO\ mission design further. For example, enhancing the sensitivity of the CMB-dominated channels by increasing the number of detectors dedicated to these frequency bands might lead to a better sensitivity in $r$ measurement.
Alternatively, we could reduce the cost of the mission by excluding these high-frequency bands.}  
\reff{In this work, we have focused on the removal of dust and hence the usefulness of high-frequency bands in the proposed design. The lower frequency bands in \ECHO\ (28-65 \GHz) are also important for the removal of the synchrotron component. It is, therefore, crucial to evaluate the effectiveness of these bands in the subtraction of complex synchrotron models. However, apart from the measurement of CMB $B$-modes, \ECHO\ has a broader science goal to measure other cosmological signals such as the thermal Sunyaev-Zeldovich (SZ) effect from galaxy clusters, CMB $E$-mode polarization, weak lensing, and possibly spectral distortion in the CMB. Hence, in order to choose an optimum set of frequency bands and coverage, it is essential to account for all the cosmological science goals and the technical constraints of the proposed instrument. We intend to investigate this in the future. }


\textbf{Note added} - Immediately prior to our paper submission to arxiv, a paper by Aurlien \textit{et.\ al.} (2022), \cite{aurlien2022foreground} appeared that also studies the impact of high-frequency bands in the recovering of $r$ for the proposed \pico\ mission. Their study also uses the \NILC\ method for the input $r_{in}=0$ case in the presence of different dust models. The key difference is that we focus our study on the \ECHO\ mission as a follow-up of our previous work \citep{10.1093/mnras/stac1474} and define the figure of merits to analyze the utility of the  high-frequency bands in recovering of the CMB $B$-mode signal. 
Aurlien \textit{et.\ al.} (2022) \cite{aurlien2022foreground} also reached a similar conclusion for the \pico\ mission that the low-frequency coverage of 21-462 GHz gives a similar level of $r/\sigma_r$ and $r_{95\%}$ (95\% confidence limit) as the high-frequency coverage of 43-799 GHz for $r_{in}=0$ and 73\% delensing case. Our main results of this paper are consistent with the results of Aurlien \textit{et.\ al.} (2022) \cite{aurlien2022foreground}.

\begin{acknowledgments}
Some of the results in this paper have been derived using the \texttt{Healpy}~\cite{zonca2019oj} and \healpix~\cite{gorski2005apj} packages. AS acknowledges the use of Padmanabha cluster\footnote{\url{https://hpc.iisertvm.ac.in/}} at IISER-TVM for her work. DA acknowledges the Department of Science and Technology, Govt. of India, for providing a Postdoctoral fellowship to contribute to this work. S.S. would like to thank NISER Bhubaneswar for the postdoctoral fellowship.  
\end{acknowledgments}



\appendix
\numberwithin{equation}{section}
\numberwithin{table}{section}
\numberwithin{figure}{section}
\section{Averaging effect}
\label{sec:averaging effect}

\begin{figure}
\includegraphics[width=\linewidth, height=0.6\textwidth]{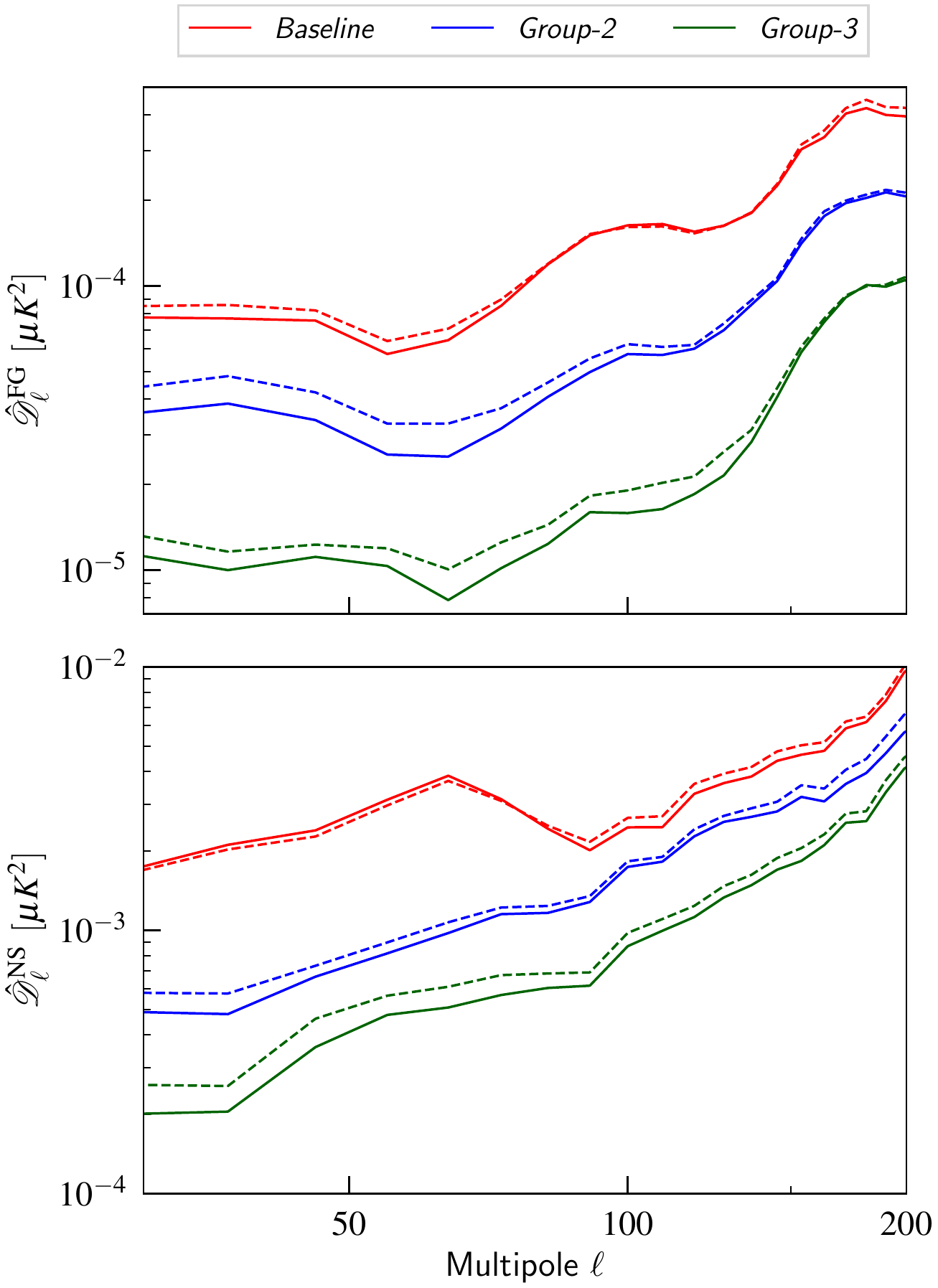}
\caption{Comparison of the level of residual foreground (\textit{upper panel}) and residual noise (\textit{lower panel}) for 3 configurations of the frequency channels. The solid lines denote the residuals for the $M2$ sky model. The dashed lines are residual levels when the beam averaging systematic effect is introduced in the $M2$-$a$ sky model. The \textit{red} color depicts the residual level for the baseline model with the least number of frequency bands in the range of 28-190 \GHz\ range. The \textit{red} and blue lines represent the residual levels for \textit{Group-2} and \textit{Group-3} configurations, respectively.}
\label{fig:d2A-2B}
\end{figure}
The SED of thermal dust emission is often modelled by an MBB power law. Since the emission law is nonlinear, it will be distorted when dust SEDs  along the LOS or instrumental beam are averaged. The SED will also be distorted if the thermal dust maps are degraded at a lower resolution by averaging over multiple pixels. This form of systematic error is unavoidable in any data analysis pipeline and may bias the estimation of $r$ ~\citep{Vacher_2022}. 
In \cite{chluba:2017}, the authors developed a moment expansion method that can be used to model SEDs with the averaging effect accurately. Since then, several component separation methods based on the moment expansion has been developed ~\citep{Rotti:2021,10.1093/mnras/stab648,Vacher_2022,10.1093/mnras/stab2392}. In  ~\cite{10.1093/mnras/stab648}, the authors have shown that \NILC\ performs less efficiently in the presence of averaging effects and moment-based semi-blind methods such as the \textit{cMILC} will give more accurate results.
\begin{table}[!htbp]
\caption{The recovered $r_{\mathrm{mp}}$ values and their $\sigma_{r}$ for the $M2$ and $M2$-$a$ sky models. The thermal dust component in the $M2$ model follows the single MBB SED. The averaging effect due to degrading the dust maps to a lower resolution has been introduced in the dust component of the $M2$-$a$ model. The fiducial \ECHO\ configuration with twenty bands is used to obtain these results.}
\label{table:D2a-result}
\begin{center}
\begin{tabular}{ccc}
\hline\hline \noalign{\vskip 2pt}
\rule[-1.5ex]{0pt}{1.ex} Sky Model &\hspace{15ex}  $r_{\mathrm{mp}} \times 10^{3}$ &\hspace{15ex}  $\sigma_{r} \times 10^{3}$ \\
\hline \noalign{\vskip 2pt}
\rule[-0ex]{0pt}{0ex}$M2$ &\hspace{15ex} -0.55 &\hspace{15ex} 0.92 \\
\rule[-0ex]{0pt}{0ex}$M2$-$a$ &\hspace{15ex} -0.59 &\hspace{15ex} 0.97 \\
\hline
\hline
\end{tabular}
\end{center}
\end{table}

Here, we test the performance of \NILC\ when systematic error due to the averaging effect is present in the dust maps. To simulate this form of modeling, we first generate the \GNILCn\ maps (described in Sec.~\ref{ssec:D2-1MBB}) at \healpix\ grid resolution of \Nside$=2048$. Next, we degrade these maps to \Nside$=256$ using \textit{map2alm} and \textit{alm2map} functions. Finally, we smooth with a Gaussian beam of FWHM=$40'$. To obtain the final sky simulations, we coadd these thermal dust maps with the simulations of CMB, synchrotron, and \ECHO\ noise. We denote this new sky model by $M2$-$a$. 
We compare the results obtained from $M2$ and $M2$-$a$. In Fig.~\ref{fig:d2A-2B}, we have shown the residuals obtained from the above two models for three selected frequency ranges. We find that for these models, residual foreground and noise levels do not differ significantly.  Hence, the $1\sigma_r$ uncertainty on the recovered $r$ and the bias on recovered $r_{\mathrm{mp}}$ (See Table~\ref{table:D2a-result}) is very similar for both cases. It is clear from these results that the performance of the \NILC\ method is not affected by the systematic errors introduced by averaging over the pixels.

\bibliography{OPT_FREQ} 

\end{document}